\documentclass[%
 reprint,
 amsmath,amssymb,
 aps,
]{revtex4-2}

\usepackage{xcolor}
\usepackage{hyperref}

\hypersetup{
    colorlinks,
    linkcolor={red!80!black},
    citecolor={blue!80!black},
    urlcolor={blue!80!black}
}
\usepackage{comment}
\usepackage{units}
\usepackage{amsmath}
\usepackage{booktabs}
\usepackage{graphicx}
\usepackage{dcolumn}
\usepackage{bm}
\usepackage{float}
\usepackage{array} 
\usepackage{booktabs} 
\usepackage [english]{babel}
\usepackage [autostyle, english = american]{csquotes}
\usepackage{multirow}

\MakeOuterQuote{"}

\graphicspath{{./}{figures/}}

\def\nar{New.~Ast.~Rev.}      
\def\apj{ApJ}                 
\def\apjl{ApJL}               
\def\apjs{ApJS}               
\def\mnras{MNRAS}             
\def\aap{A\&A}                
\def\araa{ARA\&A}             
\def\prd{Phys. Rev. D}        
\def\prl{Phys. Rev. Lett.}    

\begin{document}

\title{Self-lensing flares from black hole binaries IV: the number of detectable shadows}

\author{Kevin Park$^{1}$}
 \thanks{EMAIL: ksp2136@columbia.edu}
 \author{Chengcheng Xin$^{2}$}
\author{Jordy Davelaar$^{3,4,5,2,6}$}
\author{Zoltán Haiman$^{2,1}$}

\affiliation{
$^{1}$Department of Physics, Columbia University, New York, NY 10027, USA}
\affiliation{
$^{2}$Department of Astronomy, Columbia University, New York, NY 10027, USA}
\affiliation{$^{3}$Department of Astrophysical Sciences, Peyton Hall, Princeton University, Princeton, NJ 08544, USA
}
\affiliation{$^{4}$ NASA Hubble Fellowship Program, Einstein Fellow}
\affiliation{
$^{5}$Center for Computational Astrophysics, Flatiron Institute, 162 Fifth Avenue, New York, NY 10010, USA}
\affiliation{
$^{6}$Astrophysics Laboratory, Columbia University, 550 W 120th St, New York, NY 10027, USA}

\begin{abstract}
Sub-parsec supermassive black hole (SMBH) binaries are expected to be
common in active galactic nuclei (AGN), as a result of the
hierarchical build-up of galaxies via mergers.  While direct evidence
for these compact binaries is lacking, a few hundred candidates have
been identified, most based on the apparent periodicities of their optical
light-curves.  Since these signatures can be mimicked by AGN red-noise, additional evidence is needed to confirm their binary nature. Recurring self-lensing flares (SLF), occurring whenever the two BHs are aligned with
the line of sight within their Einstein radii, have been suggested as
additional binary signatures.  Furthermore, in many cases, lensing
flares are also predicted to contain a "dip", whenever the lensed
SMBH's shadow is comparable in angular size to the binary's Einstein
radius.  This feature would unambiguously confirm binaries and
additionally identify SMBH shadows that are spatially unresolvable by
high-resolution VLBI.  Here we estimate the number of quasars for which
these dips may be detectable by LSST, by extrapolating the quasar
luminosity function to faint magnitudes, and assuming that SMBH
binaries are randomly oriented and have mass-ratios following those in
the Illustris simulations.  Under plausible assumptions about quasar
lifetimes, binary fractions, and Eddington ratios, we expect tens of
thousands of detectable flares, of which several dozen contain
measurable dips.
\end{abstract}


\maketitle

\section{Introduction}
\label{sec:Intro}

Supermassive black holes (SMBHs) with masses between $M\approx 10^6-10^9{\rm M_\odot}$ are present in the nuclei
of most nearby galaxies \cite{Kormendy2013}. In hierarchical
cosmologies, galaxies grow by frequent mergers,
which deliver the nuclear SMBHs~\cite{Springel+2005, Robertson+2006}, along with significant quantities of gas~\cite{BarnesHernquist1992}, to
the central regions of the post-merger galaxy. The natural
conclusion is that after the two SMBH's separation decreases, SMBH binaries (SMBHBs) should form frequently in galactic nuclei \citep{begelman1980}.  Hydrodynamical simulations have shown that electromagnetic (EM) emission from these SMBHBs,
provided they are surrounded by circumbinary gas, should be detectable starting well before the merger, and should persist all the way to the merger~\citep{farris2015a,tang2018,Krauth2023,dittmann+2023}.

These compact SMBHBs are a fundamental ingredient of galaxy formation and are also prime targets to be observed in gravitational waves (GWs) by LISA~\cite{LISA2023-Pau,LISA2024-Colpi}, and by pulsar timing arrays (PTAs).   
Indeed, PTAs have recently discovered a stochastic GW background (GWB) in the nHz bands, which is consistent with the cosmological population of
coalescing SMBHBs~\cite{NANOGrav15-GWB, EPTA+InPTA-GWB, Parkes-GWB, ChinesePTA-GWB}.   
Combining the EM and GW signals from the same source - or even the same or overlapping populations of SMBHs - would open windows to especially novel science, including astrophysics, cosmology, particle physics, and the nature of gravity~\cite{baker2019}.
  
These so-called "multimessenger" opportunities have stimulated strong interest in finding wider SMBH binaries in EM data. Approximately 300 SMBH binary candidates have been identified among bright active galactic nuclei (AGN) in large time-domain optical surveys~\citep{Graham+2015b,Charisi+2016,Chen+2024} based on their apparent periodicities, and a handful of additional SMBH binary candidates have been identified serendipitously, or through other tentative signatures involving double-peaked or offset emission lines, or spatial structures of radio jets and lobes~\citep{DeRosa+2019,Bogdanovic+2022}. These candidates remain controversial because of the lack of a "smoking gun" binary signature, and in the cases of the periodic candidates, stochastic red-noise AGN variability can mimic periodicities for a few periods~\cite{Vaughan+2016}.

A potential SMBHB signature that in some cases could help lift this degeneracy is a "self-lensing flare" (SLF). If the two SMBHs are aligned within the line-of-sight to within the system's Einstein radius, then whenever one of the SMBHs passes behind the other, its emission will be strongly magnified.  These lensing flares occur once or twice per orbit (depending on whether one or both BHs are active). Depending on the binary masses and separations, the flares can last from hours to weeks~\cite{Haiman2017,dorazio2018,pihajoki2018,Ingram2021}. An AGN identified in the Kepler catalog, dubbed Spikey, has a light-curve consistent with relativistic Doppler modulation from an eccentric binary, with a narrow $\sim10\%$ spike at the expected orbital phase whose symmetric shape is well fit by a microlensing model~\citep{Hu2020}.

In a fraction of these self-lensing binaries, the size of the SMBH's shadow ($\sim5$ times its gravitational radius) is commensurate with the (angular) Einstein radius, and an additional feature is imprinted on the light-curves, in the shape of a "dip" near the peak of each flare.  This feature could unambiguously confirm binaries and
additionally identify SMBH shadows that are spatially unresolvable by high-resolution VLBI, as was done by the Event Horizon Telescope Collaboration~\citep[][hereafter Paper~I]{davelaar2022b}.   Toy models for the BH emission show that the precise shapes and sizes of these dips depend on the binary system's parameters~\citep[][Paper~II]{davelaar2022}, and recent hydrodynamical simulations find that these "dip" features exist even in the strongly distorted and fluctuating circumbinary gas, and can be recovered via phase-folding in the face of stochastic noise~\cite[][III]{Krauth_2024}. \cite{porter2024} perform a similar study of the ray-traced emission from SMBH binaries, but using a boosted binary metric and following the binary's inspiral with 3.5PN equations of motion.  They recover the self-lensing flares and dips closely matching those in Paper I.

Self-lensing flares require nearly edge-on viewing angles, and a natural question is how rare these lensed configurations are. The Vera Rubin Observatory's LSST \citep{LSST2021} is expected to contain between 20-100 million bright quasars~\citep{Xin_Haiman_2021}, and combined with its high cadence (with photometric points every few days), is an ideal dataset in which to search for rare sources with periodically recurring flares.   \citet{Kelley_2021} have recently used SMBH populations from the Illustris simulation \cite{Sijacki_2015}, combined with toy models for binary emission and lensing, and concluded that LSST could detect several hundred self-lensed SMBHs with flares lasting for 30 days or longer.

In this paper, we follow up on the above studies, to assess the number of self-lensing flares which additionally have detectable "dips" in their light-curves.    In principle, these dips require more stringent alignment, but their incidence rate rises steeply for shorter-duration flares.   Here we estimate the number of quasars for which these dips may be detectable by LSST, by extrapolating the quasar
luminosity function to faint magnitudes, and assuming that SMBH
binaries are randomly oriented and have mass ratios following those in
the Illustris simulations.  Under plausible assumptions about quasar
lifetimes, binary fractions, and Eddington ratios, we predict tens of
thousands of detectable flares with durations down to ten days, of which several dozen contain
measurable dips.

The rest of this paper is organized as follows.
In \S~\ref{sec:Methods}, we describe our methodology, including models for the SMBH binary populations, their emission and lensing, the criteria for detectable flares and dips, and the impact of finite source sizes.
In \S~\ref{sec:DisCon}, we present our results, in terms of the number of flares and dips detectable in LSST as a function of binary parameters and observational thresholds (magnitude and flare duration).
In \S~\ref{sec:DisCon}, we summarize our main results and their implications.

\section{Methods}
\label{sec:Methods}

In this section, we calculate the number of detectable self-lensing dips in LSST and its dependence on several binary parameters. 
Results from \cite{Kauffmann_Haehnelt_2000} demonstrate that the evolution of quasars can be reproduced in a model in which they are activated in galaxy mergers.  As mentioned above, these galaxy mergers are expected to deliver the two SMBHs to the new galactic nucleus where they form a bound binary~\cite{begelman1980}.  
Given these results, our main assumption is that galaxy mergers are responsible for both quasar activity and for producing SMBH binaries~\cite{Haiman_Kocsis_Menou_2009}. First, we find the expected number of quasars ($N_{\rm QSO}$) in LSST using the quasar luminosity function (\S~\ref{subsec:m_z}). We then modify $N_{\rm QSO}$ to find the number of binary quasars as a function of the binary orbital period, using estimates of the quasar lifetime at each orbital period (\S~\ref{subsec:t_orb}) allowing the overall binary fraction to be a free parameter. We find the mass ratio distribution for these binaries in different mass and redshift bins using the Illustris simulations (\S~\ref{subsec:q}) and we compute the probability of a detectable self-lensing dip given binary inclination (\S~\ref{subsec:phi}) as a function of the mass, mass ratio and orbital period. Finally, we compute similar probabilities for self-lensing flares in the point-source limit (\S~\ref{subsec:point-equations}) and accounting for the finite sizes of the emitting regions (\S~\ref{subsec:finite-equations}).

\subsection{Number of quasars and binaries} \label{subsec:m_z}
We follow the calculations in \cite{Xin_Haiman_2021} to obtain the number of quasars above LSST's detection threshold, based on the extrapolated quasar luminosity function \cite[QLF;][]{Kulkarni_Worseck_Hennawi_2019b}. 
\begin{figure}
    \centering
    \includegraphics[width=1\linewidth]{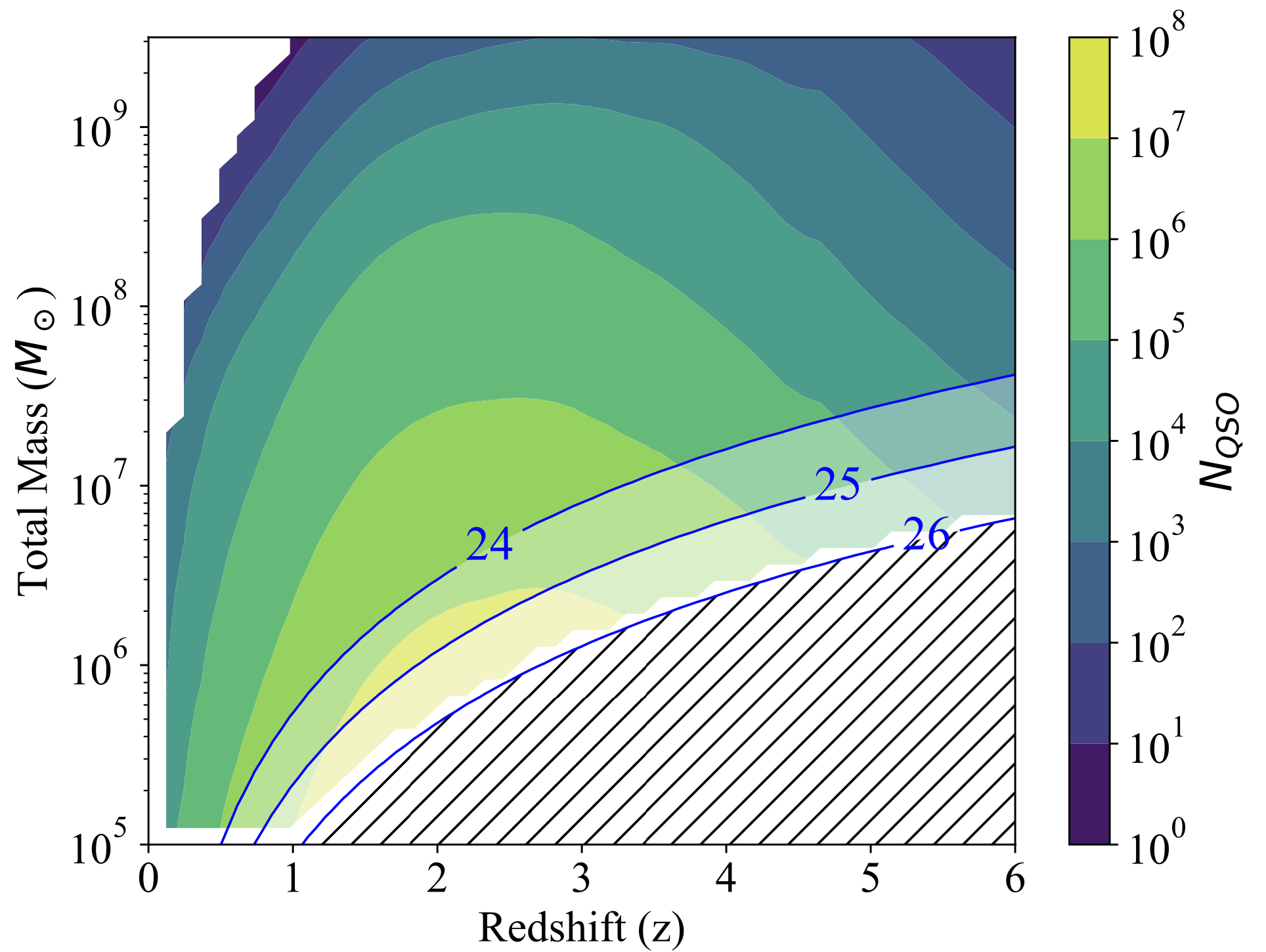}
    \caption{The number of quasars $N_{\rm QSO}(M,z)$ in the redshift and mass range $z\in [0,6]$ and $M/{\rm M_\odot}\in[10^5,10^{9.5}]$. Integrating over the full mass and redshift ranges gives 20 million, 44 million, 100 million quasars above the assumed LSST magnitude detection limits of $m_i=24,25,26$, in agreement with the results of \cite{Xin_Haiman_2021}.}
    \label{fig:QLF}
\end{figure}
In Figure~\ref{fig:QLF}, we present the number of quasars in LSST's 20,000 deg$^2$ survey, $N_{\rm QSO}$, for three magnitude limits, $m_i=24$, $25$ and $26$, where $m_i=24$ corresponds to the single-exposure magnitude limit of LSST in the $i$ band, and $m_i=26$ corresponds to the co-added magnitude limit over the whole survey \cite{LSST_2009}. The BH mass and $i$-band magnitudes are related using Eq.~\ref{eq:mi_mbh}:
 \begin{equation}\label{eq:mi_mbh}
        m_i = 24+2.5\log\left[\left(\frac{f_{\rm Edd}}{0.3}\right)^{-1}\left(\frac{M}{3\cdot 10^6 M_\odot}\right)^{-1}\left(\frac{d_L(z)}{d_L(z=2)}\right)^2\right],
\end{equation} 
where 
$f_{\rm Edd}$
is defined by the bolometric quasar luminosity $L=f_{\rm Edd}L_{\rm Edd}$ and $L_{\rm Edd}$ is the Eddington luminosity for total mass $M$. Initially, we assume $f_{\rm Edd}=0.3,$ but in \S~\ref{sec:DisCon} below we will discuss the dependence of our results to varying $f_{\rm Edd}$. Quasars in the hatched area are discarded since their magnitudes are below $m_i=26$.

Under the assumption that all quasars are triggered by mergers and are associated with SMBH binaries \cite{Sanders_1990, Hopkins_2008}, and that the bright quasar phase lasts for a typical lifetime of $\tau_Q$ (=say $10^8$ years), the number of quasars powered by SMBH binaries with $\tau_m$ years left to merger scale linearly with the fraction $\tau_m/\tau_Q$, 
\begin{equation}
    N(M,z)\equiv \left( \frac{\tau_m}{\tau_q}\right) f_{\rm bin} N_{\rm QSO}(M,z).
\end{equation} 
$N_{\rm QSO}(M,z)$ is the number of quasars shown in Figure~\ref{fig:QLF}, which assumes that all quasars are associated with binaries. In this analysis, we vary the fraction of quasars that are SMBH binaries, $f_{\rm bin}$, from 0.2 to 1. We use residence times $\tau_m=\tau_{\rm res}(q,M,T_{\rm orb},z)$ appropriate to GW- or gas-driven binary inspirals, depending on the mass $M$ and orbital period $T_{\rm orb}$ of the binary (see next section). By setting a maximum period $T_{\rm max}$, $N(M,z)$ represents the number of binaries with an orbital period of $T_{\rm max}$ or less, equivalent to the number of binaries with $\tau_m$ years left to merger. 

\begin{figure}[t]
    \centering
    \includegraphics[width=1\linewidth]{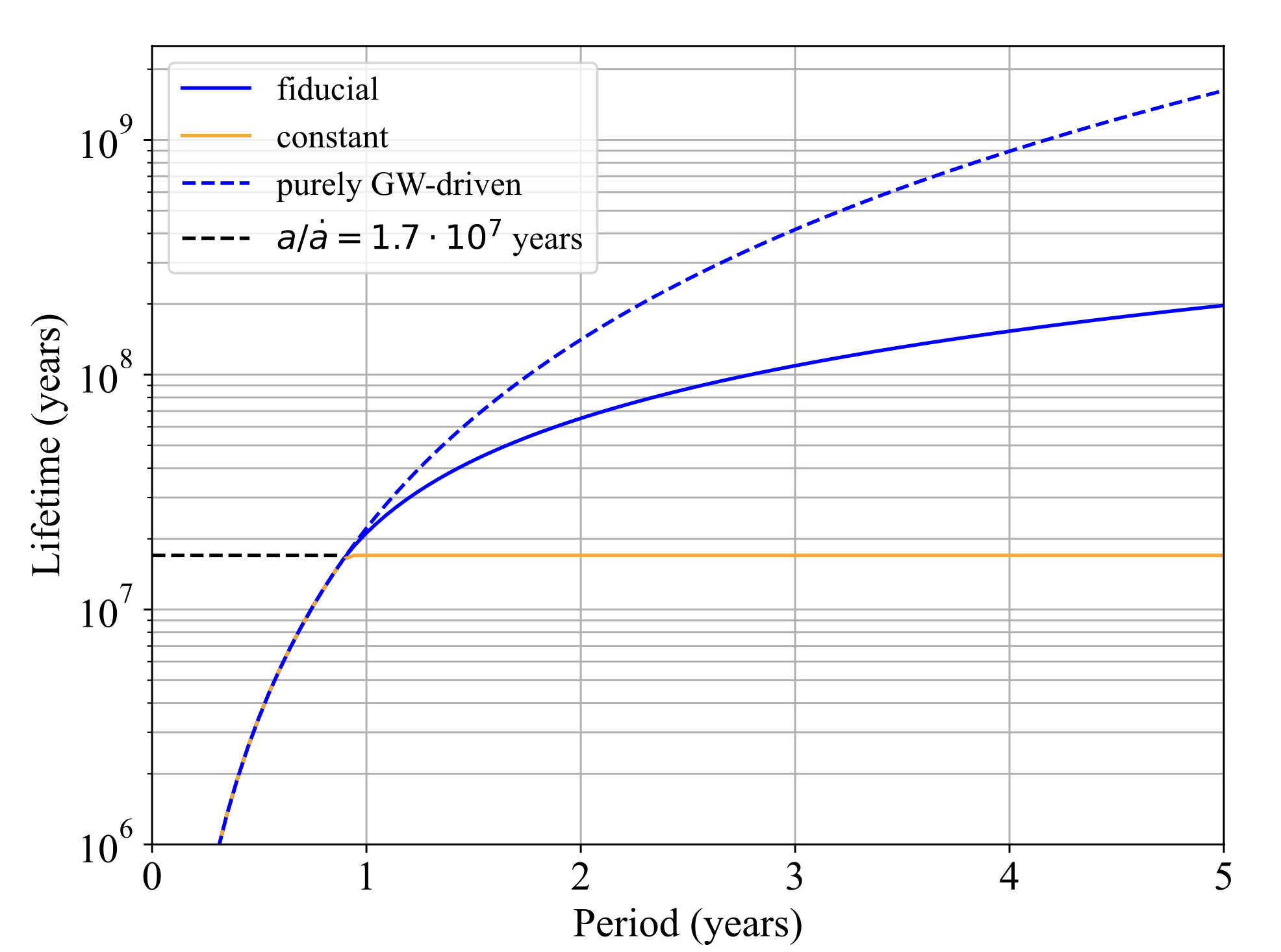}
    \caption{The solid blue curve shows the fiducial model for the residence time $\tau_{\rm res}(q,M,T_{\rm orb},z)$ of binaries with total mass $M=10^{7}M_\odot$,  mass ratio $q=1$ and 
    redshift $z=1$ as a function of observed orbital period $T_{\rm orb}$. Below the threshold given in Eq. \ref{eq:5} the binaries evolve via GW emission and above the threshold the binaries' inspiral timescale is assumed to be driven by circumbinary gas and increase linearly with $T_{\rm orb}$.
    In \S~\ref{sec:DisCon} we also present results for a more conservative model where a constant maximum lifetime of $a/\dot{a}_{\rm gas}=1.7 \times 10^7$ years is imposed (Eq.\ref{eq:6}), shown by the solid orange curve. Purely GW-driven inspiral, the dashed blue curve, is shown for reference.}
    \label{fig:lifetime}
\end{figure}

\subsection{Residence time vs. orbital period}  \label{subsec:t_orb}
We obtain the residence time of a binary quasar at a given orbital period (or separation) $\tau_{\rm res}$ as follows. For short periods, the orbital decay is primarily GW-driven, given by the quadrupole formula \cite{peters1964}:
\begin{equation} \label{eq:3}
    \tau_{\text{GW}}=1.11\times 10^7  q_s^{-1}M_7^{-5/3}\left(\frac{T_{\rm orb}}{\text{yr}}\right)^{8/3}\left(\frac{1+z}{4}\right)~\text{yr},
\end{equation}
where $q_s \equiv 4q/(1+q)^2$ is the symmetric mass ratio, $q\equiv M_2/M_1\leq 1$ is the mass ratio, $M_7 \equiv M/(10^7 M_\odot)$ is the total mass in units of $10^7$ solar masses, and $T_{\rm orb}$ is the orbital period in units of years.

\begin{figure*}[t]
    \centering
    \includegraphics[width=1\linewidth]{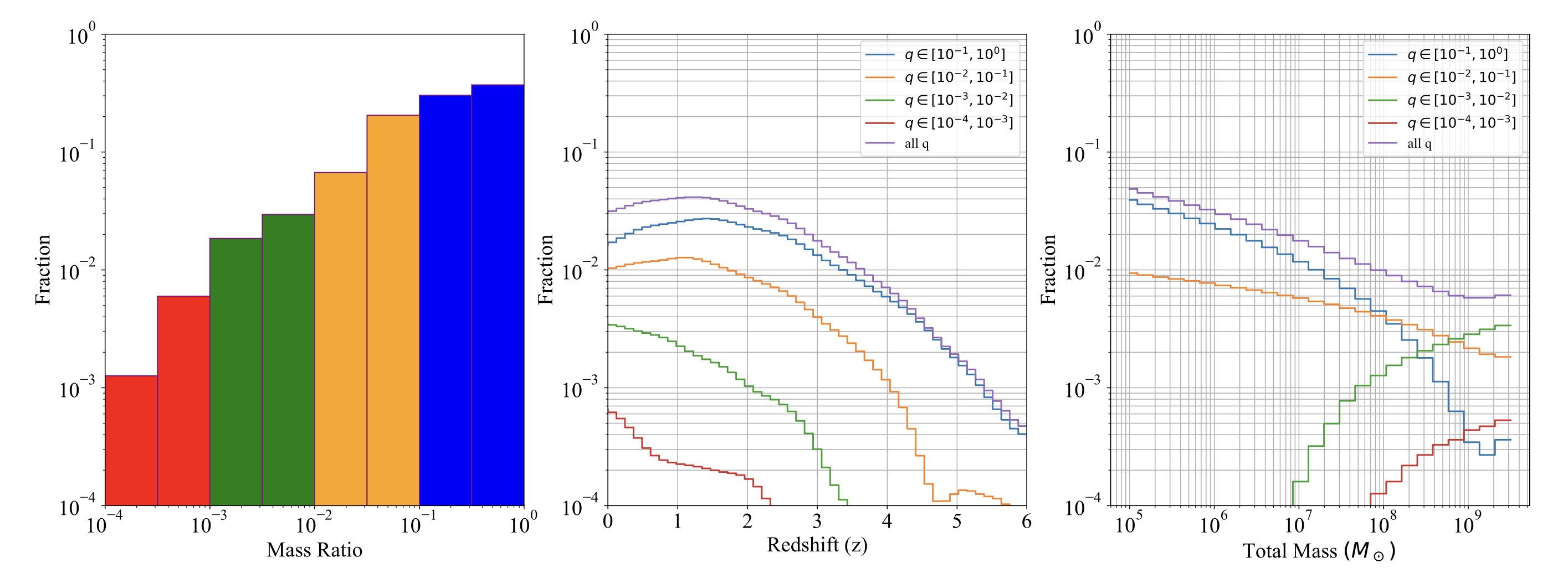}
    \caption{The left panel shows the distribution of the mass ratios of SMBH binaries in the Illustris simulations. In the middle and right panels, we illustrate the redshift and total mass distributions of Illustris binaries, in the four different mass-ratio ranges shown by colors in the left panel. Most binaries are symmetric in mass ($q>0.1$) and have relatively low redshift ($z<3$) and total mass ($M<10^7 M_\odot)$.}
    \label{fig:Illustris}
\end{figure*}

For long periods, we assume that the circumbinary gas dominates the binary's inspiral, and adopt the simple scaling from \cite{Bortolas_2021}. This assumes that the gas-driven inspiral timescale is proportional to the mass accretion timescale, roughly in agreement with the results of hydrodynamical simulations \cite{Tang+2017,Muñoz_2019, Moody+2019,Muñoz_2020,duffell2020,Tiede+2020}. First, the mass accretion timescale is referenced in units of the Eddington rate:
\begin{equation}
\begin{split}
        \dot{m}_{\rm Edd}=\frac{4\pi G m m_p}{\eta \sigma_T c}= 2.26 \cdot 10^{-2}\times \left(\frac{\eta}{0.1}\right)^{-1}\\ \times \left(\frac{m}{10^6 M_\odot}\right)M_\odot \text{yr}^{-1},
\end{split}
\end{equation} where $G$ is the gravitational constant, $m$ is the mass of the accreting BH, $m_p$ is the proton mass, $c$ is the speed of light, $\sigma_T$ is the Thomson scattering cross section and $\eta$ is the radiative efficiency, assumed to be $0.1$. Assuming an Eddington accretion rate $\dot{m}=\dot{m}_{\text{Edd}}$, then 
\begin{equation} \label{eq:5}
    \frac{m}{\dot{m}}\sim 4.4 \times 10^7 \text{yr},
\end{equation} 
and we follow \cite{Bortolas_2021} and adopt the residence time at binary separation $a$
\begin{equation}\label{eq:6}
    \tau_{\text{gas}}=\frac{a}{\dot{a}_{\text{gas}}}=\frac{1}{2.68}\frac{m}{\dot{m}} \sim 1.7 \times 10^7 \text{yr},
\end{equation}
which is constant (independent of $a$).
At large separations for which the GW-inspiral timescale exceeds this threshold, we assume as our fiducial model that the binary gas-driven residence time increases linearly with the orbital period, as shown by the solid blue curve in Figure~\ref{fig:lifetime}. As an alternative model, the residence time is fixed at a maximum of $1.7 \times 10^7$ years, shown by the solid orange curve. The illustrations in Figure~\ref{fig:lifetime} are for a binary with mass $M=10^7{\rm M_{\odot}}$, redshift $z=1$ and mass ratio $q=1$. Finally, we constrain the maximum observed orbital period to $T_{\text{max}}=5$ years, given that for LSST's survey of 10 years, we want to observe a few recurrences of the dip.

\subsection{Mass-ratio distributions}
\label{subsec:q}
The GW-driven inspiral timescale in Eq.~\ref{eq:3} depends on the binary's mass ratio $q$, which is not constrained by the QLF or other observations. To estimate the mass-ratio distribution we instead use Illustris-3 \cite{Sijacki_2015}, a cosmological hydrodynamical simulation that self-consistently follows the evolution and mergers of galaxies and their central SMBHs. Illustris provides the merger tree of their SMBH mergers, and we use this to compute to distribution of $q$ as a function of total mass $M$ and redshift $z$, shown in Figure~\ref{fig:Illustris}. Illustris embeds MBHs of seed mass $\sim 10^5 M_\odot$ which accrete and evolve dynamically. Due to the limitations to spatially resolve closely separated low-mass binaries ($M<10^6{\rm M_{\odot}}$), we initially implement a conservative mass cut of $10^6 M_\odot$ of Illustris BH binaries and extrapolate the count distribution in $(M,z)$ down to the binary total mass of $10^5 M_\odot$. We then divide the Illustris binaries by their mass ratios into 8 logarithmic bins between $10^{-4}\leq q\leq 1$. Figure~\ref{fig:Illustris} depicts the fraction of Illustris binaries in each of these 8 bins (left panel). The colors blue, yellow, green, and red each correspond to the binaries with mass ratios in the ranges of $[10^{-1}, 1], [10^{-2}, 10^{-1}], [10^{-3}, 10^{-2}], [10^{-4}, 10^{-3}]$.
Using Illustris, we calculate the probabilities that a BH binary with given $M$ and $z$ in one of 50 bins has a mass ratio $q$ in one of the 8 logarithmic bins--see the middle and right panel of Figure~\ref{fig:Illustris}. Using these mass ratio distributions we can evaluate the GW-driven inspiral timescale for binary quasars in the QLF.

\subsection{Self-lensing dips} \label{subsec:phi}
As mentioned in \S~\ref{sec:Intro}, for SMBHBs with nearly edge-on orbital planes, periodic self-lensing flares occur as a result of gravitational lensing. Compact binaries close to merger, with an orbital period of 5 years or less in our fiducial model (Table \ref{table_fiducial}),
are expected to have four or more self-lensing flares detectable within a full ten-year LSST survey. Additionally, GR ray tracing simulations of~\cite{davelaar2022b} reveal that self-lensing flares have observable dips, caused by the black hole shadow. In this work, we estimate the number of LSST binaries with detectable self-lensing dips for different binary parameters. For this, we recap Eqs.~1-3 of \cite{davelaar2022b}, which give an analytical expression for the probability that a binary system with given binary mass ratio $q$, total mass $M$, orbital period $T_{\rm orb}$ and redshift $z$ has a detectable self-lensing dip.

First, assuming a circular binary with nearly edge-on binary inclination, the expected phase spacing between the two peaks before and after the dip is the ratio of the diameter of the BH shadow and the circumference of the orbit: 
\begin{equation}\label{eq:phase_dip}
    \Delta \phi = \frac{d_{\text{shadow}}}{2\pi a_{\text{orb}}},
\end{equation}
where $a_{\text{orb}}$ is the orbital radius, $d_{\text{shadow}}=2\sqrt{27}GM_{\text{source}}/{c^2}$ is the BH shadow diameter of the source, and $M_{\text{source}}=qM/(1+q)$ is the mass of the lensed BH (assumed here to be the lower-mass secondary). Expressing the orbital radius in terms of binary parameters via $a_{\text{orb}}=(GMT_{\rm orb}^2/4\pi^2)^{1/3}$ gives
\begin{equation}
    \Delta \phi = 5.63 \frac{q M^{2/3}}{(1+q)T_{\rm orb}^{2/3}} \, {\rm radians},
\end{equation} 
where $G=c=1$ units were used. $T_{\rm orb}$ has units $R_{\rm g}/c$, where $R_{\rm g}=GM_{\text{source}}/c^2$ is the gravitational radius of $M_{\text{source}}$. 
The dip in the self-lensing flare can be observed for a range of binary inclinations, $\Delta i$. When $\Delta i$ is smaller than the angular size of the BH shadow on the sky, i.e. when Eq.~\ref{eq:9} is satisfied, the focal point of the lens will be eclipsed by the BH shadow, causing dips at the local maxima of the flares. The inclination window for which this dip is visible is given by
\begin{equation}\label{eq:9}
    \Delta i \leq  \sin^{-1}(\pi \Delta \phi) \equiv \Delta i_{\rm dip}.
\end{equation}
Assuming that BH binary orbital inclinations are randomly distributed on a unit sphere, the minimum angular separation between the BHs on the sky occurs when the two BHs are at either end of the semi-minor axes of the orbit's projected ellipse. Then, for a binary chosen at random, the probability that it has an observable self-lensing dip during its orbit is given by 

\begin{equation} \label{eq:13}
    P_{\rm dip}=\Delta i_{\rm dip}(q,T_{\rm orb},M,z)/90^{\circ},
\end{equation}
i.e. the probability that $\Delta i$, which can range from $0^{\circ}$ to $90^{\circ}$, satisfies Eq.~\ref{eq:9}. The number of detectable self-lensing dips is then $N_{\rm dips}=P_{\rm dip}\times N(M,z)$, integrated appropriately over the distribution of all binary parameters of mass ratio, observed orbital period, total mass and redshift.

\begin{table}[t]
\begin{tabular}{@{}lllll@{}}
    \toprule
    Parameter  & Fiducial                 & Range                      &  &  \\ \midrule
    $f_{\rm bin}$    & 1                        & $0.2-1$                      &  &  \\
    $f_{\rm Edd}$    & 0.3                      & $0.1-0.5 $                   &  &  \\
    $T_{\rm max}$     & 5 yrs                     & $2.5-10$ yrs                  &  &  \\
    $\tau_Q$     & $10^8$ yrs & $10^{7-9}$ yrs &  &  \\
    $\tau_{\rm res}$   & Figure \ref{fig:lifetime}, blue solid         & Figure \ref{fig:lifetime}, orange solid         &  &  \\
    $\tau_{\rm f,min}$ & 0 days                   & $0-30$ days   \\
    \bottomrule
\end{tabular}
\caption{Parameters in our fiducial model (top row) and their ranges considered (bottom row). $f_{\rm bin}$ is the fraction of quasars associated with binaries, $f_{\rm Edd}$ is their Eddington ratio, $T_{\rm max}$ is the maximum orbital period of interest, $\tau_Q$ is the average total bright quasar lifetime, $\tau_{\rm res}$ is the residence time, i.e. the duration a binary quasar spends at each orbital period (shown in Fig.~\ref{fig:lifetime}), and $\tau_{f,{\rm min}}$ is the minimum required lensing flare duration (not imposed in the fiducial model).}
\label{table_fiducial}
\end{table}

\subsection{Self-lensing flares - point source limit}
\label{subsec:point-equations}
We wish to compare our calculations with \cite{Kelley_2021}, who have calculated the number of detectable self-lensing flares, irrespective of whether dips from the BH shadow are measurable, based on the abundance of binaries in Illustris. To do this, we construct a self-lensing probability similar to the self-lensing dip probability in Eq.~\ref{eq:13}. We adopt the point-source (PS) magnification \cite{Paczynski_1986b} \begin{equation}
    \label{eq:point-source mag}
    \mathcal{M}_{\rm PS} = \frac{u^2+2}{u\sqrt{u^2+4}},
\end{equation} where $u=\text{Re}(u_1+u_2)$ is the projected separation in units of the Einstein radius and \begin{equation}
    u_j = \left[\frac{ac^2(\cos^2\phi_j+\sin^2i\sin^2\phi_i)}{4G(M-m_j)\cos i\sin\phi_j}\right]^{1/2},
\end{equation} where $i$ is again the binary inclination relative to the line of sight and $\phi_j$ is the orbital phase of each BH. At the peak of the self-lensing flare ($\phi_2 = \pi/2$), we assume the secondary BH to be the source, and find the maximum orbital inclination $\Delta i_{\rm PS}$ in which a 10\% magnification occurs, i.e. $\mathcal{M}_{\rm PS}(i)>1.1$. The corresponding self-lensing probability is \begin{equation}
    P_{\rm PS}=\Delta i_{\rm PS}(q,T_{\rm orb},M,z) / 90^\circ
\end{equation} and the number of self-lensing flares in the point-source limit is then $N_{\rm PS} = P_{\rm PS}\times N(M,z)$, again integrated appropriately over the distribution of all binary parameters.

\subsection{Self-lensing flares - finite source}
\label{subsec:finite-equations}

According to Figure 2 of \cite{D’Orazio_Di_Stefano_2017}, for binaries of total mass $\lesssim 10^8 M_\odot$ and orbital periods of a few years, it is necessary to incorporate the accretion disc size of the secondary because the angular size of the accretion disc becomes comparable to the Einstein radius for lower total masses. We adopt their model to account for finite-source effects, based on a multi-color accretion disc extending from the innermost stable circular orbit (ISCO) $r_{\rm ISCO} = 6GM_{\rm source}/c^2,$ to the secondary's tidal truncation radius $r_{\rm tidal}=0.27 q^{0.3} a$ \cite{Roedig_Krolik_Miller_2014}
The temperature profile of the disc is given by \begin{equation}
    \label{T(r) function}
    \sigma T^4(r)=\frac{3GM_{\rm source}\dot{M}_{\rm source}}{8\pi r^3}\left[ 1-\left(\frac{r_{\rm ISCO}}{r}\right)^{1/2}\right],
\end{equation} 
where $r$ is the distance from the central SMBH and $\dot{M}_{\rm source}$ is the accretion rate of the lensed BH, for which we assume an Eddington accretion rate. For $r>r_{\rm tidal}$ and $r<r_{\rm ISCO},$ we set $T(r)=0.$ The resulting flux from the disc is given by \begin{equation} \label{eq:black-body}
    F_\nu (r) = \pi B_\nu [T(r)],
\end{equation}
where $B_\nu$ is the Planck function. We calculate the lensing magnification by evaluating \begin{equation}
    \label{finite-source magnification}
    \mathcal{M}_{\nu}^{\rm FS} = \frac{\int_0^{2\pi}\int_0^{\infty}F_\nu(u',v')\mathcal{M}_{\rm PS}(u')u'du'dv'}{\int_0^{2\pi}\int_0^{\infty}F_\nu(u',v')u'du'dv'},
\end{equation}
where $\mathcal{M}_{\rm PS}(u')$ is given in Eq. \ref{eq:point-source mag}, evaluated in the lens-centered polar coordinates $(u,v)$: \begin{equation}
    r_*^2=(u_0^2+u^2-2u_0u\cos(v-v_0))r_E^2
\end{equation}
\begin{equation}
    r=r_*\sqrt{\cos^2\phi + \frac{\sin^2\phi}{\cos^2(\pi/2-J)}}
\end{equation}
\begin{equation}
    \sin \phi = \frac{u\sin v - u_0 \sin v_0}{\sqrt{(u\sin v - u_0 \sin v_0)^2+(u\cos v - u_0 \cos v_0)^2}}.
\end{equation}
Here $(u_0, v_0)$ is the position of the secondary, and $J$ is the inclination of the source disc relative to the line of sight. We define $\Delta i_{FS}$ as the maximum orbital inclination such that $\mathcal{M}_\nu^{\rm FS}(i)>1.1$ at alignment and the probability that a binary has a detectable self-lensing flare in the finite source limit is 
\begin{equation}\label{eq:finite-source probability}
    P_{\rm FS} = \Delta i_{\rm FS}(q,T_{\rm orb},M,z,\nu,J)/90^\circ. 
\end{equation} 
The number of self-lensing flares in the finite-source limit is $N_{\rm FS}=P_{\rm FS}\times N(M,z)$, integrated over binary parameters. 

In this paper, we consider the representative observed wavelengths near the center of the six LSST filters $u, g, r, i, z, y$ at 380, 476, 622, 755, 870, 1015 nm, which are chosen by taking the average of the FWHM transmission points of each filter \cite{LSST_2009}. For example, for a binary at redshift $z\in [0,6]$, the rest-frame wavelength of $380/(1+z)$ nm will contribute flux to the u-band, and the rest-frame wavelength is used to evaluate the wavelength-dependent flux in Eq. \ref{eq:black-body}. 

\vspace{3mm} 

Absorption by hydrogen clouds in the intergalactic medium could attenuate quasar light depending on the observing wavelength and the quasar redshift. However, according to \cite{Madau_1995}, for observed wavelengths of $\nu\sim 400$ nm, the mean cosmic transmission is close to 1 for $z\lesssim 3$, where the majority of quasars are. For longer observed wavelengths of $\nu> 700$ nm, the mean cosmic transmission is close to 1 for $z\lesssim 5$, which means intergalactic absorption is mostly negligible for our purposes. In all calculations, we fix the source-disc inclination at a representative value of $J=\pi/4$ for simplicity.

\begin{figure}
    \centering
    \includegraphics[width=1\linewidth]{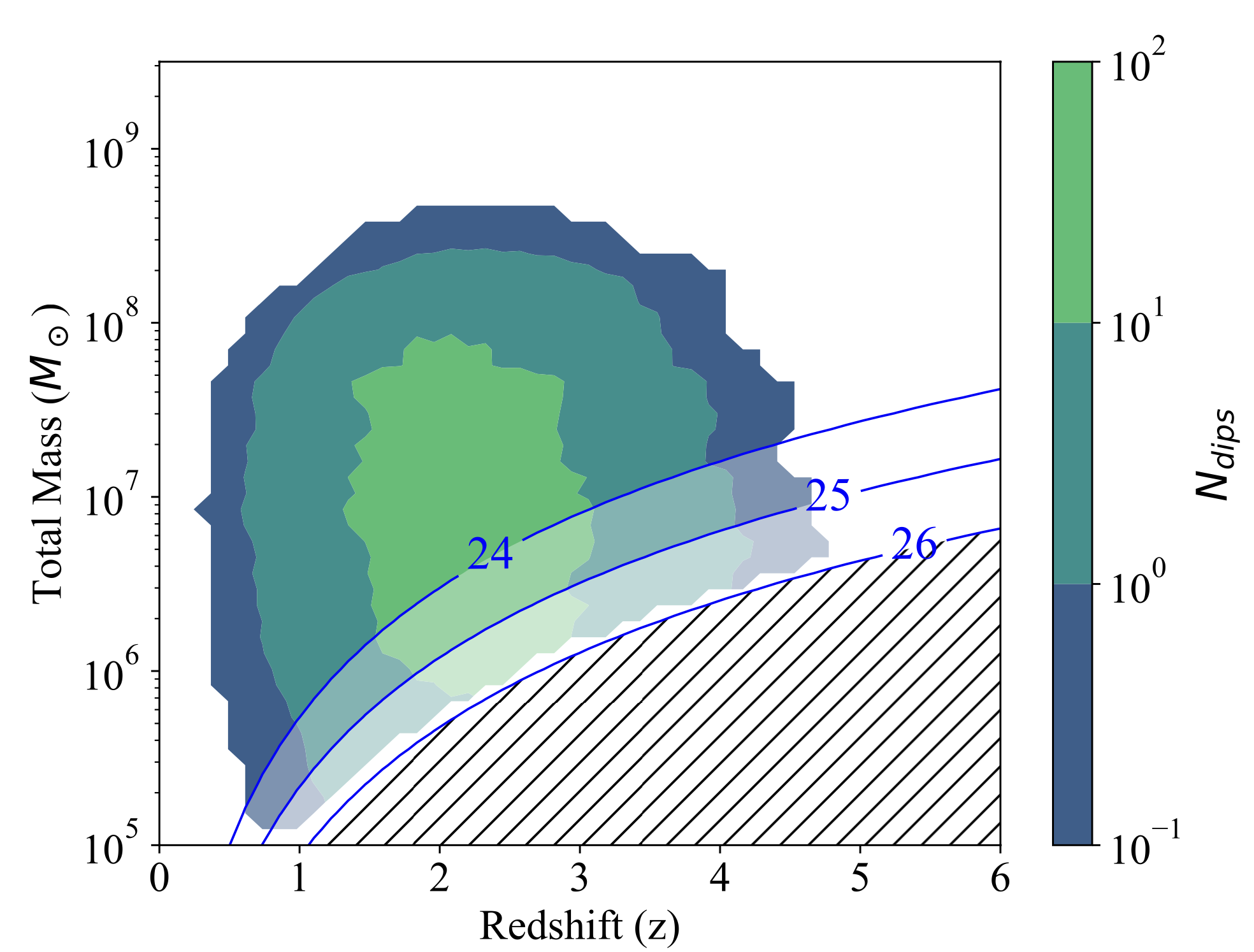}
    \caption{The distribution of the number of detectable self-lensing dips in the $M-z$ plane. The contours for $N_{\rm dips}$ are in units of per unit redshift per unit log mass bin, where the unit bin sizes are $\Delta z = 0.12$ and $\Delta \log_{\rm 10} M = 0.08$. Three LSST sensitivity limits, covering the range from a single exposure to the fully co-added survey detection threshold, are shown for reference. Most detectable dips are from quasars at $1.5\lesssim z \lesssim 3$ with total (binary) SMBH masses of $10^6 M_{\odot}\lesssim M \lesssim 10^8 M_{\odot}$.}
    \label{fig:N_dips}
\end{figure}

\section{Results \& discussion}
\label{sec:Results}
\subsection{Self-lensing dips and their mass- and redshift-dependence}
We first present the number of detectable dips, for total SMBH binary masses and redshifts in the ranges of $M/{\rm M_\odot}\in [10^5, 10^{9.5}]$ and $z\in [0, 6]$. These mass- and redshift-distributions in the fiducial model are shown in Figure~\ref{fig:N_dips}, and are also presented numerically in Tables~\ref{tab:M-dep} and \ref{tab:z-dep}.  Overall, we find 41-60 detectable dips, depending on the magnitude threshold.
Table~\ref{tab:M-dep} shows that $\sim 90\%$ of these binaries with detectable dips have masses between $10^6 M_{\odot}$ to $10^8 M_{\odot}$. Qualitatively, while the distribution of quasars (Fig.~\ref{fig:QLF}) is concentrated at low masses (below $10^6 M_\odot$), the self-lensing probability in Eq.~\ref{eq:13} increases with mass, resulting in the distribution clustering in the intermediate mass regime, see Table~\ref{tab:M-dep}.

\vspace{3mm} 

Table~\ref{tab:z-dep} shows that approximately $80\%$ of these binaries are between redshifts $z=1-3$, which is also expected considering that the lensing probability depends only weakly on redshift and the QLF peaks in this range.

\vspace{3mm} 

\begin{table}
    \begin{tabular}{cccc}
        \toprule
        Mass ($M_\odot$) & \textbf{$m_i=24$} & \textbf{$m_i=25$} & \textbf{$m_i=26$} \\
        \midrule
        $[10^{5}, 10^{6}]$  & 0  & 1  & 3 \\
        $[10^{6}, 10^{7}]$  & 12 & 21 & 28 \\
        $[10^{7}, 10^{8}]$  & 25 & 26 & 26 \\
        $[10^{8}, 10^{9}]$  & 4  & 3  & 3 \\
        $[10^{9}, 10^{10}]$ & 0   & 0   & 0 \\
        \midrule
        \textbf{Total:}     & \textbf{41} & \textbf{51} & \textbf{60} \\
        \bottomrule
    \end{tabular}
    \caption{Dependence of self-lensing dips on binary mass, integrated over all redshifts $z\in [0,6]$ and mass ratios $q\in [10^{-4}, 1]$. This result also assumes the fiducial parameters in Table \ref{table_fiducial}.}
    \label{tab:M-dep}
\end{table}
\begin{table}
    \centering
    \begin{tabular}{cccc}
        \toprule
        Redshift ($z$) & \textbf{$m_i=24$} & \textbf{$m_i=25$} & \textbf{$m_i=26$} \\
        \midrule
        $[0,1]$  & 1  & 1  & 1 \\
        $[1,2]$  & 16 & 18 & 20 \\
        $[2,3]$  & 19 & 24 & 29 \\
        $[3,4]$  & 5  & 7  & 9 \\
        $[4,5]$ & 0   & 1   & 1 \\
        $[5,6]$ & 0   & 0   & 0 \\      
        \midrule
        \textbf{Total:}     & \textbf{41} & \textbf{51} & \textbf{60} \\
        \bottomrule
    \end{tabular}
    \caption{Redshift-dependence of self-lensing dips, integrated over all masses $M/ M_\odot\in [10^{5},10^{9.5}]$ and mass ratios $q\in [10^{-4}, 1]$. The conditions for $f_{\rm bin}$ and lensing duration were applied as for the total mass dependence in Table~\ref{table_fiducial}.}
    \label{tab:z-dep}
\end{table}
Furthermore, the LSST limits on apparent magnitude, $m_i<24$ (or, optimistically, $m_i=26$) constrain the $M-z$ parameter space, which is visualized in the contours of Figure~\ref{fig:QLF}. Therefore, we present all of our results below for three different LSST magnitude limits, $m_i=24$, 25 and 26, covering the range from a single exposure limit to the fully co-added survey detection threshold. Figure~\ref{fig:N_dips} also shows these three magnitude cuts (blue curves), with quasars below $m_i=26$ are discarded as too faint to be detected (hatched region).
\subsection{Mass-ratio dependence}
\begin{table}
    \centering
    \begin{tabular}{cccc}
        \toprule
        Mass ratio & \textbf{$m_i=24$} & \textbf{$m_i=25$} & \textbf{$m_i=26$} \\
        \midrule
        $[10^{-0.5}, 10^{0.0}]$   & 33   & 41   &   50    \\
        $[10^{-1.0}, 10^{-0.5}]$  & 7    & 9    &   9     \\
        $[10^{-1.5}, 10^{-1.0}]$  & 1     & 1     &   1      \\
        $[10^{-2.0}, 10^{-1.5}]$  & 0     & 0     &   0      \\
        $[10^{-2.5}, 10^{-2.0}]$  & 0     & 0     &   0      \\
        $[10^{-3.0}, 10^{-2.5}]$  & 0     & 0     &   0      \\
        $[10^{-3.5}, 10^{-3.0}]$  & 0     & 0     &   0      \\
        $[10^{-4.0}, 10^{-3.5}]$  & 0     & 0     &   0      \\
        \midrule
        \textbf{Total:}      & \textbf{41} & \textbf{51} & \textbf{60} \\
        \bottomrule
    \end{tabular}
    \caption{Mass-ratio dependence of self-lensing dips, integrated over all redshifts $z\in [0,6]$ and total masses $M\in [10^{5}, 10^{9.5}] M_\odot$. This result also assumes the fiducial parameters in Table \ref{table_fiducial}.}
    \label{tab:q-dep}
\end{table}

There are no ab initio constraints on the Illustris mass ratios, but in practice we find all of them to be within $q=10^{-4} - 1$, with almost all ($>98\%$) binaries in the mass-ratio range of 0.1 to 1, as shown in Table \ref{tab:q-dep}. This follows from the mass ratio distribution of SMBH binaries from Illustris--the left panel of Figure \ref{fig:Illustris}. 
We note that these data from Illustris are quite uncertain, especially since all of the SMBH binary merger physics is sub-grid. In particular, mergers are assumed to take place instantly when their separation (modeled semi-analytically) decreases below a certain smoothing length. In reality, there can be delays and disruptions by a third SMBH, which can change the mass-ratio distribution \cite{Sayeb_2023}, and in turn the number of self-lensing dips in each $q$-bin. We leave a full exploration of this dependence to future work.

\subsection{Dependence on model parameters}

\begin{table}
    \centering
    \begin{tabular}{ccccccc}
        \toprule
         $f_{\text{bin}}$ & $f_{\text{\rm Edd}}$ & $T_{\text{max}}$ (yr) & $\tau_Q$ (yr) & $m_i=24$ & \textbf{$m_i=25$} & $m_i=26$\\
        \midrule
        \vspace{0.3cm}
        1   & 0.3 & 5 & $\tau_Q$ & 41 & 51 & 60 \\

        \textbf{0.8} & 0.3 & 5 & $10^8$ & 33 & 41 & 48 \\
        \textbf{0.6} & 0.3 & 5 & $10^8$ & 25 & 31 & 36 \\
        \textbf{0.4} & 0.3 & 5 & $10^8$ & 16 & 20 & 24 \\
        \vspace{0.3cm}
        \textbf{0.2} & 0.3 & 5 & $10^8$ &  8 &  10 & 12 \\
        1   & \textbf{0.5} & 5 & $10^8$ & 27 & 34 & 40 \\
        \vspace{0.3cm}
        1   & \textbf{0.1} & 5 & $10^8$ & 72 & 101 & 132 \\
        1   & 0.3 & \textbf{10} & $10^8$ & 63 & 76 & 87 \\
        \vspace{0.3cm}
        1   & 0.3 & \textbf{2.5} & $10^8$ & 24 & 32 & 40 \\
        1   & 0.3 & 5 & $\mathbf{10^9}$ & 4 & 5 & 6 \\
        1   & 0.3 & 5 & $\mathbf{10^7}$ & 412 & 503 & 601 \\
        \bottomrule
    \end{tabular}
    \caption{Number of detectable dips varying with model parameters and LSST limits: $f_{\text{bin}}$ is the assumed fraction of quasars that are binaries, $f_{\text{\rm Edd}}$ is the Eddington ratio, $T_{\text{max}}$ is the maximum allowed orbital period, and $\tau_Q$ is the typical bright quasar lifetime. The first row corresponds to our fiducial model, while each subsequent row varies one of the parameters, shown in bold. Various maximum allowed lensing durations $\tau_{f,{\rm min}}$ were additionally considered and are shown separately in Table~\ref{table_lensdur}.}
    \label{tab:param-dep}
\end{table}

We next examine how the number of detectable dips depends on the orbital period and other model parameters.  The dependence on $T_{\rm max}$ comes from the probability of an observable self-lensing dip being proportional to $\Delta \phi \propto T_{\rm orb}^{-2/3}$, 
and the residence time as a function of period $\tau_{\rm res} \propto T_{\rm orb}^{8/3}$ at shorter orbital periods and $\tau_{\rm res} \propto T_{\rm orb}$ at longer orbital periods. Doubling the fiducial maximum period to 10 years would yield an increase in the number of self-lensing dips by a factor between 4 and 1.26.
This is consistent with the increase by a factor of $\sim1.5$ as shown in Table~\ref{tab:param-dep} when $T_{\rm max}=10$ years. 

Our results also depend on the assumed Eddington ratio $f_{\rm Edd}$, average quasar lifetime $\tau_Q$ and the fraction of quasars that are binaries, $f_{\rm bin}$, as presented in Table~\ref{tab:param-dep}.

According to Eq.~\ref{eq:mi_mbh}, changing $f_{\rm Edd}$ from $0.3$ to $0.1$ would increase the total mass of our BHs by a factor of~3. The total mass dependence in our model also comes from the self-lensing dip probability $\propto M^{4/3},$ and the residence time, which is either $\propto M^{-5/3}$ for GW-driven inspiral or independent of total mass ($\propto M^0$) for gas-driven inspiral. Therefore, we expect decreasing $f_{\rm Edd}$ from $0.3$ to $0.1$ to increase the number of detectable dips by a factor between 0.7 and 4.3, which is consistent with our results, where we find a factor of $\sim 2$ increase. 

Finally, our results are linearly proportional to $\tau_Q^{-1}$ and $f_{\rm bin}$, which causes the number of dips to vary within a few orders of magnitude.
As mentioned, initially we assume that all quasars are binaries ($f_{\rm bin}=1$). In reality, the fraction of quasars that are binaries is a function of binary parameters $M$ and $z$. In Table~\ref{tab:param-dep}, we list the numbers for varying $f_{\rm bin}$. The numbers are also sensitive to the model for the residence time $\tau_{\rm res}$. We find that if we conservatively set a maximum residence time of $1.7\times 10^7$ years (orange curve in Figure \ref{fig:lifetime}), then there are only $5,6,7$ self-lensing dips for the limits $m_i=24,25,26$. We also note that the expected phase spacing of the dip, Eq. \ref{eq:phase_dip}, is valid only for circular orbits and would not hold for non-circular orbits.

\subsection{Self-lensing flares and comparison to K21}
\label{subsec:point-results}

\begin{figure}
    \centering
    \includegraphics[width=1\linewidth]{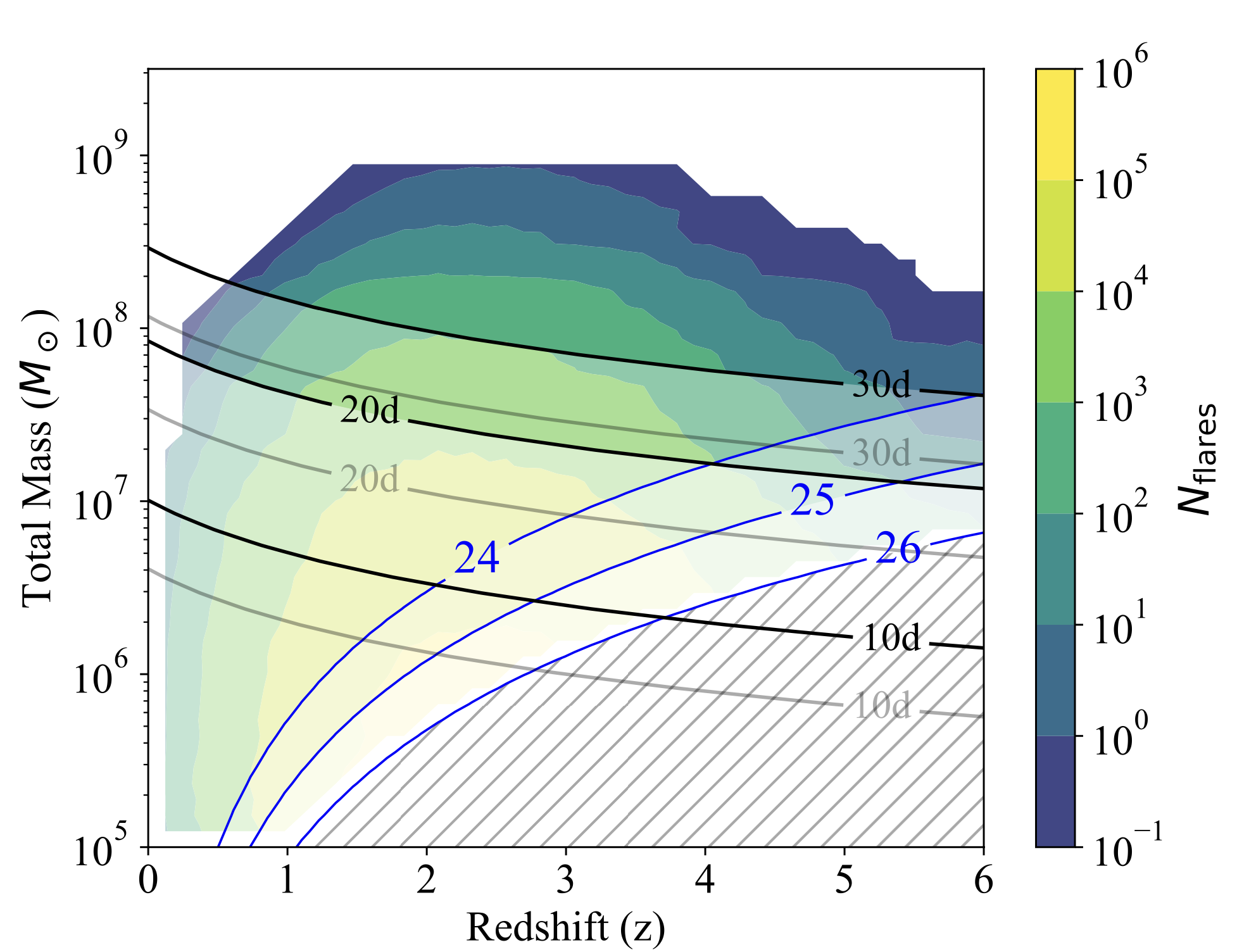}
    \caption{Distribution of self-lensing flares in the point-source approximation, without requiring dips. Blue curves show LSST sensitivity limits, as usual, and the black contour lines show cutoffs resulting from requiring that lensing durations exceed 30, 20, and 10 days for our typical binaries with $q=1, T_{\rm orb}=5$ years. The gray contour lines correspond to the same minimum lensing durations but for binaries with $q=0.1$, which are the typical binary parameters when requiring a minimum lensing duration of 30 days. Changing the mass ratio from $q=1$ to $0.1$ increases $M_{\rm lens}$ by a factor of $\sim 2$ and lower total masses are needed to produce the same lensing duration contour. We further demonstrate the change in the mass-ratio distribution before and after the lensing requirement in Figure~\ref{fig:q distribution}.}    \label{fig:n_flares}
\end{figure}

Although our main focus in this paper is on the number of detectable dips, we next discuss the number of detectable lensing flares, without considering whether or not a measurable dip may be present. As mentioned in the Introduction, the number of detectable flares has been previously estimated by \cite{Kelley_2021}. We therefore make some adjustments to our fiducial model, to mimic the assumptions in K21 as closely as we can, and then compare our results to theirs in detail.
The results, analogous to the number of dips, are shown in Figure~\ref{fig:n_flares}.

\begin{table}
    \centering
    \begin{tabular}{ccccccc}
        \toprule
        $\tau_{\rm f, min}$ & dip/flare & $m_i=24$ & $m_i=25$ & $m_i=26$\\
        \midrule
        0d & dip & 41 & 51 & 60 \\
        \textbf{10d} & dip & 36 & 45 & 48 \\
        \textbf{20d} & dip & 23 & 24 & 24 \\
        \vspace{0.3cm}
        \textbf{30d} & dip & 9 & 9 & 9 \\
        0d & \textbf{flare (PS)} & 44037 & 97118 & 195754 \\
        \textbf{10d} & \textbf{flare (PS)} & 21242 & 32473 & 42143 \\
        \textbf{20d} & \textbf{flare (PS)} & 4083 & 4415 & 4484 \\
        \textbf{30d} & \textbf{flare (PS)} & 734 & 737 & 737 \\
        \bottomrule
    \end{tabular}
    \caption{The number of detectable self-lensing dips and flares as a function of the required minimum lensing duration $\tau_{\rm f, min}$ and LSST magnitude limit. The point-source (PS) approximation is assumed for the flares. We show results for various LSST sensitivity limits. Likewise, the first row shows the results in our fiducial model, with the parameters that are varied shown in bold in subsequent rows. The number of detectable flares is a steep function of the minimum required flare duration.  
    Figures~\ref{fig:n_flares} and \ref{fig:3-panels} (right panel) visualize the effects of requiring these lensing durations.}
    \label{table_lensdur}
\end{table}

To make a direct comparison with the results of \cite{Kelley_2021}, we need to apply similar constraints on our SMBH population from the QLF, in terms of lensing amplification, sensitivity limits, lensing durations and orbital periods. We therefore adopt K21's sensitivity limit of $3\times 10^{-30}\; \text{erg}\; \text{s}^{-1} \text{cm}^{-2} \text{Hz}^{-1}$, which corresponds to $m_i=25.2$. We also require that self-lensing durations exceed 30 days (10 intra-flare data points assuming three-day LSST cadence) using the Einstein radius at alignment 
\begin{equation}
    r_{\rm E} = \sqrt{2ar_{\rm s} \cos I_{\rm orb}},
\end{equation} 
where $r_{\rm s}=2GM_{\rm lens}/c^2$ is the Schwarzschild radius of the lens, $M_{\rm lens}=M/(1+q)$. The approximation for the lensing duration is given by
\begin{equation} \label{eq: lensing duration}
    \tau_f = \frac{T_{\rm orb}}{\pi}\sin^{-1}\left(\frac{r_{\rm E}}{a}\right),
\end{equation} 
where $T_{\rm orb}$ is the observed orbital period. Finally, we consider only binaries with an observed orbital period of $T_{\rm max}\leq 5$ years. Evaluating the number of detectable self-lensing flares in our BH population gives 737 detectable self-lensing flares, compared to their $450^{+29}_{-40}$ flares, where the superscript and subscript denote the interquartile ranges for their expected number. See Table \ref{tableVI} for a detailed summary of the two models, where we compare the parameter ranges of the BH populations, the SLF requirements, and the medians of various binary parameters of the SLFs. 

\begin{figure}
    \centering
    \includegraphics[width=1\linewidth]{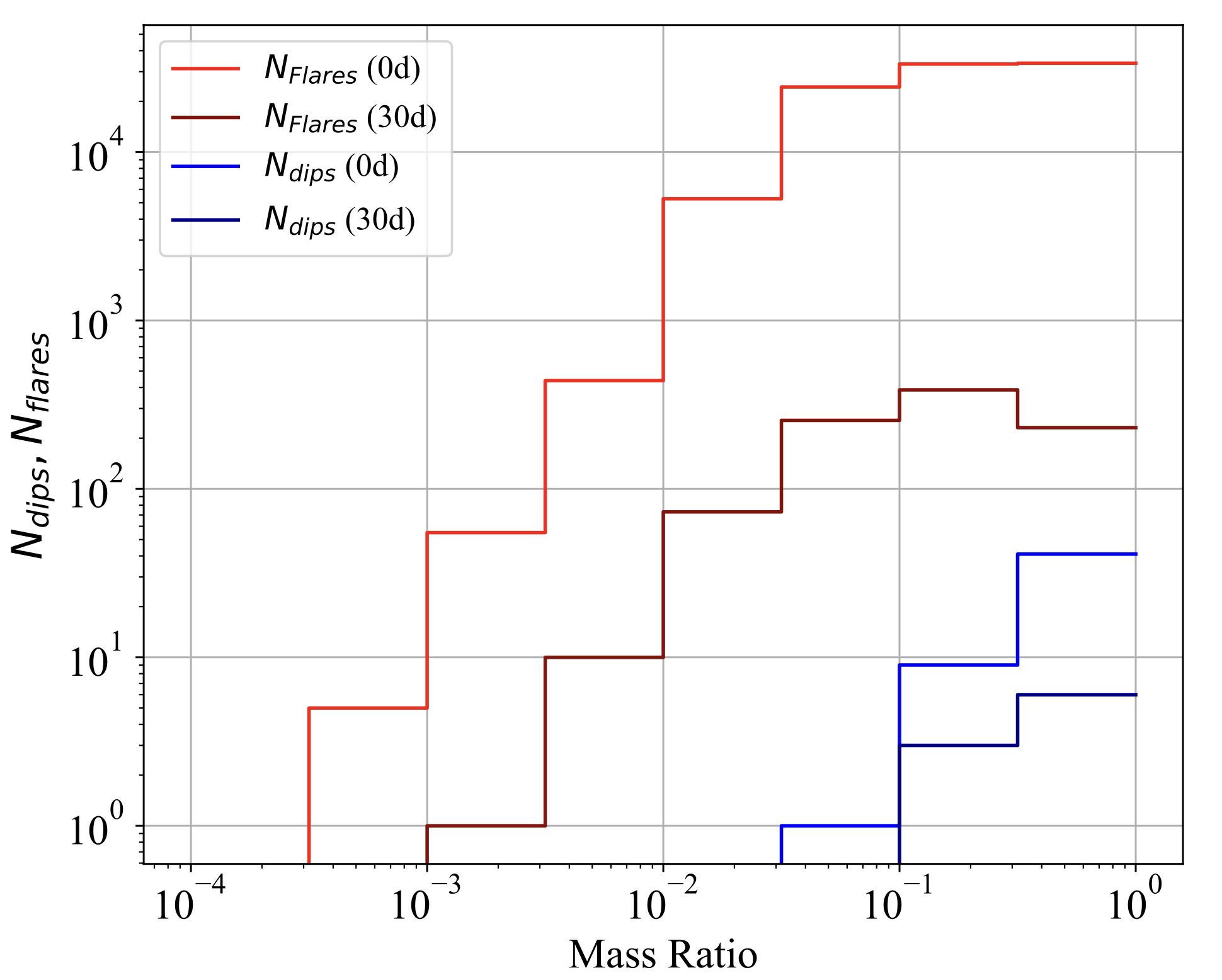}
    \caption{The mass-ratio distribution of dips and flares in the point-source approximation, where a minimum lensing duration of 0 days or 30 days is required. The two blue $N_{\rm dips}$ distributions correspond to the mass ratio distributions of the 51 and 9 dips given in the $m_i=25$ column of Table \ref{table_lensdur}. The two red $N_{\rm flares}$ distributions correspond to the 97118 and 737 flares of the same $m_i=25$ column in Table \ref{table_lensdur}.}
    \label{fig:q distribution}
\end{figure}

In both models, it is notable that as a result of requiring a minimum lensing duration of 30 days, the median  mass of the binaries with SLFs are
above $\sim10^{8} M_\odot$, significantly higher than the typical mass of both parent SMBH  populations, $10^{6-7} M_\odot$.  This occurs because the lensing duration increases with total mass. Furthermore, the lensing duration is longer when, for a given total mass $M$, the mass of the primary $M_{\rm lens}=M/(1+q)$ is larger. As a result, both models select relatively asymmetric mass ratios ($q=10^{-2}\sim 10^{-1}$) from the parent SMBH population, which is predominantly symmetric ($q=1$). Increasing the orbital period increases the lensing duration and so binaries with orbital periods of several years tend to exhibit detectable self-lensing flares. Finally, the redshift dependence of the various lensing requirements is weak and the median redshift is mostly unchanged relative to the parent SMBH binary population.

\begin{table*}
    \centering
    \begin{tabular}{l|c|c|c}
        \toprule
        & Parameter & This work & Kelley et al. (2021) \\
        \midrule
        \multirow{6}{*}{Parameter ranges of SMBHs} & Total mass $M/M_\odot$ & $[10^5, 10^{9.5}]$ & $[10^6, 10^{10}]$ \\
        & Redshift $z$ & [0,6] & $[10^{-1}, 10^1]$ \\
        & Mass ratio $q$ & $[10^{-4},10^{0}]$ & $[10^{-4},10^{0}]$ \\
        & Eddington ratio $f_{\rm Edd}$ & 0.3 (fixed) & $[10^{-5},10^{0}]$ \\
        & Orbital period $T_{\rm orb}$ & [0,5] years & [0,10] years 
        \\
        & Total \# of BHs & $4.4 \times 10^7$ & $1.5 \times 10^6$ \\
        \midrule
        \multirow{4}{*}{Flare requirements} & Magnification $\mathcal{M}_{\rm PS}$ & $>1.1$ & $>1.05 + 1/SNR$ \\
        & Min. Lensing Duration $\tau_{\rm f, min}$ & $>30$ days & $>30$ days \\
        & Max. orbital period & $<5$ years & $<5$ years \\
        & Sensitivity limit $m_i$ & $<25$ & $<25.2$ \\
        \midrule
        \multirow{4}{*}{Median of SLF parameters} & Total mass $M$ & $2.2\times 10^8 M_\odot$ & $4.4\times 10^8 M_\odot$ \\
        & Redshift $z$ & 2.1 & 0.75 \\
        & Mass ratio $q$ & 0.1 & $0.035$ \\
        & Orbital period $T_{\rm orb}$ & 4.0 years & 3.7 years \\
        \midrule
        \multicolumn{2}{c|}{\textbf{Total \# of SLFs}} & \textbf{737} & \textbf{450} \\
        \bottomrule
    \end{tabular}
    \caption{
    Comparisons between our model {\it vs.} Kelley et~al.~\cite{Kelley_2021}. The {\em top part} of the table compares our assumed parent SMBH populations, where we rely on the QLFs and use Illustris only for mass ratios. In contrast, \cite{Kelley_2021} self-consistently evolve $\sim 10^4$ SMBH binaries extracted from Illustris using a semi-analytical model and a re-sampling scheme. Our QLF-based population extends to 10 times lower masses and has roughly 30 times more BHs. We also fix the Eddington ratio to 0.3, whereas \cite{Kelley_2021}'s binaries have varying Eddington ratios typically at $\sim 0.01$. For a binary with given total mass $M$ and redshift $z$, our binaries are $\sim 30$ times brighter than K21's binaries, which could account for why we predict $2-10$ times more detectable flares for minimum lensing durations of $10-30$ days. In the {\em middle part}, we compare the requirements imposed on BH binaries for detectable self-lensing flares. The only significant difference is in the minimum magnification, where we require $10\%$, whereas \cite{Kelley_2021} require a minimum $5\%$ magnification in addition to the inverse of signal-to-noise ratio $SNR$, which is dependent on their damped random walk model for intrinsic AGN variability. Finally, in the {\it bottom part}, we compare the median parameters of binaries that exhibit detectable self-lensing flares and also the final number of self-lensing flares. Despite our different approaches, the predicted number of flares are comparable (737 {\it vs.} 450).}
    \label{tableVI}
\end{table*}

While K21 requires a minimum flare duration of 30 days, equivalent to 10 intra-flare data points, it could be advantageous to include the much more numerous shorter flares (e.g. 10 or 20 days for 3 or 6 intra-flare data points) and search for additional data in higher-cadence surveys or perform targeted high-cadence follow-up observations.
We find that there is a steep dependence on the lensing duration, where $\sim10^{4-5}$ flares are reduced to just 734-737 after requiring the lensing duration of 30 days. We show the dependence of the number of flares on their duration in Table~\ref{table_lensdur}. We fix all other parameters to their fiducial values (first row of Table~\ref{table_fiducial}). In Figure~\ref{fig:n_flares}, we further show the cut on the SLF binaries due to minimum duration requirements for two different mass ratios, $q=1$ (black curves) and $q=0.1$ (gray curves). 

Finally, we demonstrate the mass-ratio distribution of the self-lensing flares and dips in our model in Figure~\ref{fig:q distribution}. The mass-ratio distributions of flares and dips without requiring a minimum lensing duration are shown in red and blue. We assume $m_i=25$ and the other fiducial parameters given in Table \ref{table_fiducial}. In this case, most flares and dips have $q=1$, following the mass-ratio distribution of Illustris binaries, previously shown in Figure~\ref{fig:Illustris}. The mass-ratio distributions of flares and dips after requiring a minimum 30-day lensing duration are shown in dark red and dark blue. As a result of the minimum lensing-duration requirements,
the peak of the distribution shifts to $q=0.1$, while the qualitative trend of $N_{\rm dips}$ is relatively unchanged because there is only 1 dip for $q<0.1$. 

\subsection{Self-lensing flares - finite source sizes}
\label{subsec:finite-results}

\begin{figure}
    \centering
    \includegraphics[width=1\linewidth]{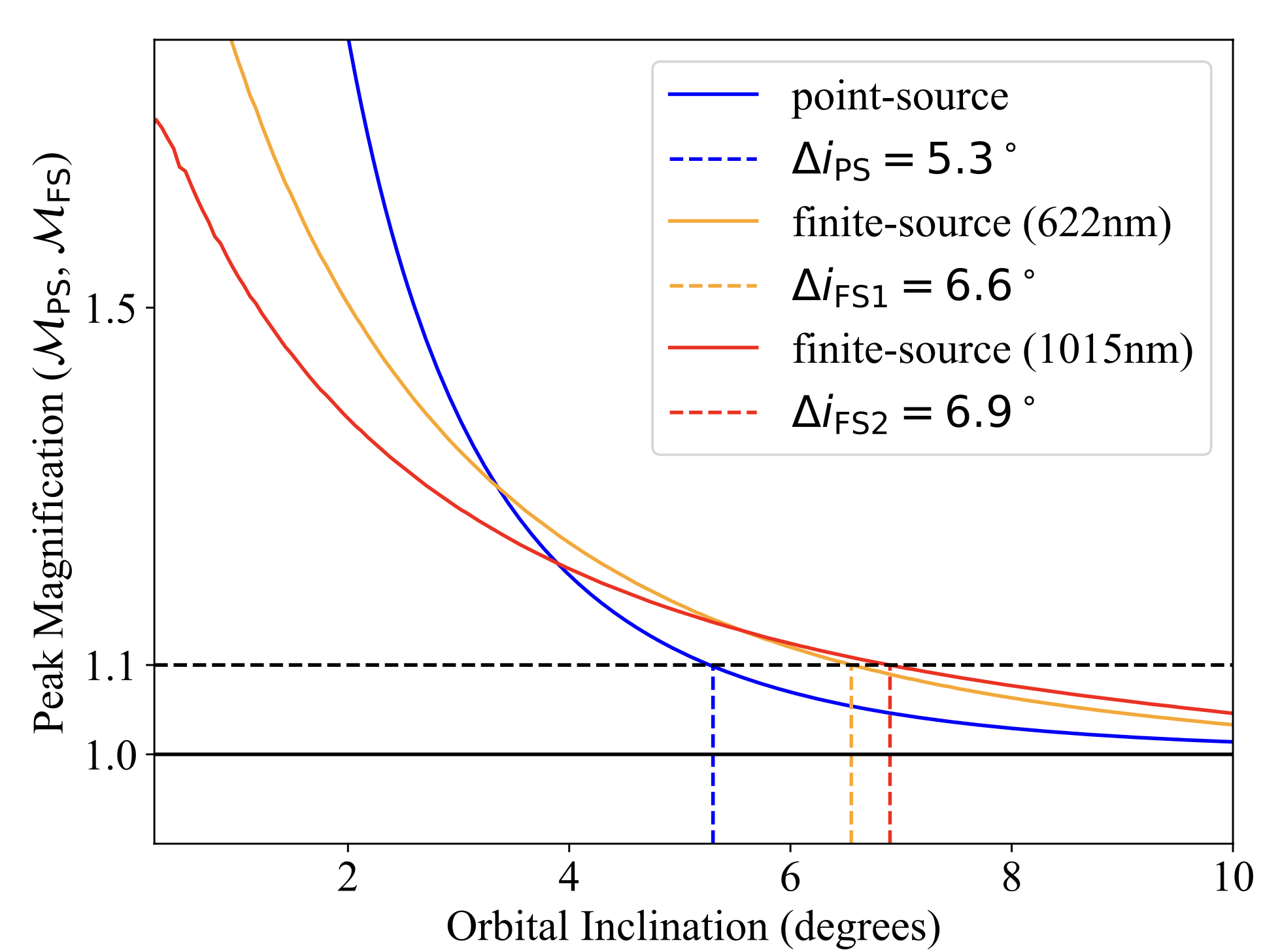}
    \caption{Peak magnification for a binary with $q=1, M=10^8 M_\odot, z=2, T_{\rm orb}=5$ years. In the finite-source approximation, peak magnifications in the r-band (622 nm) and y-band (1015 nm) are shown as a function of orbital inclination in degrees. In this sample binary, the largest orbital inclination which admits a $10\%$ flare is $\Delta i_{\rm FS1}/\Delta i_{\rm PS}=1.24$ times larger for the finite-source flare in the r-band compared to point-source flare. This occurs for most of the binaries in our BH population. Between the 622 nm flare and the 1015 nm flare, there is only a $\Delta i_{\rm FS2}/\Delta i_{\rm FS1}=1.05$ factor difference, which accounts for the mild increase in the number of detectable flares for longer wavelengths. For reference, for this binary $\Delta i_{\rm dip}=0.1^\circ,$ which is $50-70$ times smaller than $\Delta i_{\rm PS}$ or $\Delta i_{\rm FS}.$}
    \label{fig: PS vs FS}
\end{figure}

To predict the number of detectable self-lensing flares in the previous section, we used the analytic methods in \S~\ref{subsec:point-equations}, which assume that the background (lensed) SMBH is a point source. However, in some cases, the size of the source, assuming the emission arises from a minidisk modeled as a standard $\alpha$-disk, becomes comparable to the Einstein radius, and must be taken into account~\cite{D’Orazio_Di_Stefano_2017}.  

To see how a finite source size impacts our results, we re-compute the number of detectable lensing flares using the equations from \S~\ref{subsec:finite-equations}.   Overall, we find a $\sim 25\%$ increase in the number of detectable self-lensing flares relative to the point-source approximation, across the various wavelengths and LSST sensitivity limits, which we summarize in Table~\ref{table_wavelength_dependence}.

\begin{table}[H]
    \centering
    \begin{tabular}{ccccccc}
        \toprule
        LSST Filter / & & & & \\
        wavelength & $\tau_{\rm f, min}$ & $m_i=24$ & $m_i=25$ & $m_i=26$\\ \hline
        \multirow{2}{*}{u-band /} & 0d & 63113 & 144458 & 301393 \\ & 10d & 27092 & 41338 & 53595 \\
        \multirow{2}{*}{340 nm} & 20d & 5156 & 5569 & 5653 \\ & 30d & 919 & 923 & 924 \\ \hline
         
        \multirow{2}{*}{g-band /} & 0d & 64457 & 147527 & 307490 \\ & 10d & 27658 & 42137 & 54509 \\
        \multirow{2}{*}{476 nm} & 20d & 5242 & 5656 & 5741 \\ & 30d & 931 & 936 & 936\\ \hline

        \multirow{2}{*}{r-band /} & 0d & 65918 & 150598 & 313926 \\ & 10d & 28449 & 43253 & 55839 \\
        \multirow{2}{*}{622 nm} & 20d & 5377 & 5794 & 5878 \\ & 30d & 950 & 955 & 955 \\ \hline
        
        \multirow{2}{*}{i-band /} & 0d & 66707 & 152163 & 316995 \\ & 10d & 29036 & 44115 & 56902 \\
        \multirow{2}{*}{755 nm} & 20d & 5484 & 5905 & 5990 \\ & 30d & 965 & 969 & 969 \\ \hline
        
        \multirow{2}{*}{z-band /} & 0d & 67071 & 152778 & 317867 \\ & 10d & 29410 & 44709 & 57655 \\
        \multirow{2}{*}{870 nm} & 20d & 5571 & 5995 & 6080 \\ & 30d & 976 & 980 & 980\\ \hline
        
        \multirow{2}{*}{y-band /} & 0d & 67284 & 152816 & 317718 \\ & 10d & 29792 & 45294 & 58430 \\
        \multirow{2}{*}{1015 nm} & 20d & 5661 & 6090 & 6175 \\ & 30d & 987 & 991 & 991 \\ 
        \bottomrule
    \end{tabular}
    \caption{The number of detectable self-lensing flares in the finite-source approximation for different required minimum lensing durations $\tau_{\rm f, min}$ in different LSST filters/wavelengths. The three right-most columns show results for various assumed detection limits.}
    \label{table_wavelength_dependence}
\end{table}

This result is somewhat counter-intuitive, since naively, one would expect point sources to be more highly magnified.   To understand why there are more detectable flares from finite-sized sources, 
we consider $\Delta i_{\rm PS}$ and $\Delta i_{\rm FS}$, which are defined by the largest orbital inclination that produces a self-lensing flare of $10\%$. These are proportional to the probabilities $P_{\rm PS}, P_{\rm FS}$ that a binary exhibits a detectable self-lensing flare.
We visualize the differences between $\Delta i_{\rm PS}$ and $\Delta i_{\rm FS}$ in Figure \ref{fig: PS vs FS}.  This figure shows that
for orbital inclinations close to zero, the point-source magnification at alignment is much greater than the finite-source magnification, but the opposite happens for larger orbital inclinations. For the typical binaries that constitute most of our BH population with $q\sim 1, M/M_\odot \in [10^6, 10^8], z\in [1,3], T_{\rm orb}\sim 5$ years, we find that $\Delta i_{\rm FS}/\Delta i_{\rm PS}=1.1-1.3$, which accounts for the difference between the number of point-source flares and the number of finite-source flares. Notably, for less-common binaries with $q<0.1$ or $M>10^9 M_\odot$, we find that the point-source and finite-source magnifications converge, i.e. $\Delta i_{\rm PS}\sim \Delta i_{\rm FS}$, and these sources do not affect the overall difference in the number of flares. 

\begin{figure*}
    \centering
    \includegraphics[width=1\linewidth]{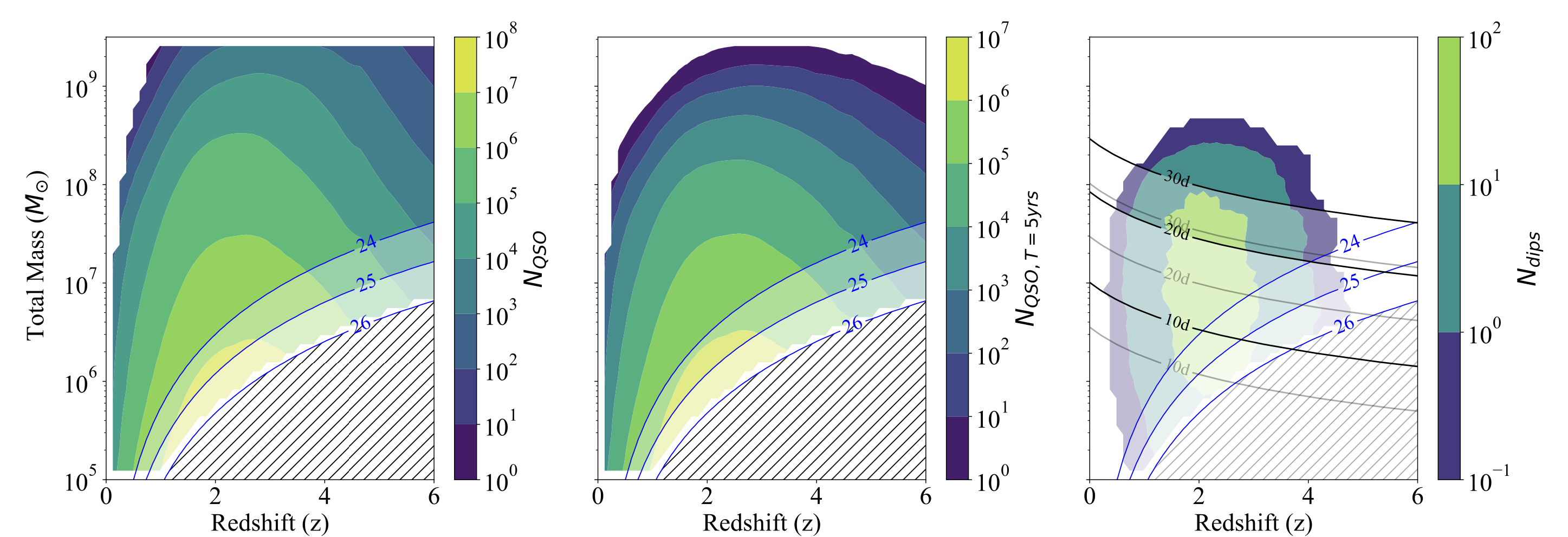}
    \caption{Illustration of going from the parent SMBH population to the detectable lensing dips. {\it Left panel}: Mass- and redshift-distribution of all quasars (same as Fig.~\ref{fig:QLF}). For the LSST sensitivity limits of $m_i=24, 25, 26$, we find 20 million, 44 million, and 100 million quasars. {\it Middle panel}: Number of binaries with an orbital period of 5 years or less. For the LSST sensitivity limits of $m_i=24,25,26$, we find 7 million, 17 million, and 45 million binaries. {\it Right panel}: Number of detectable self-lensing dips in binaries (same as Fig.~\ref{fig:N_dips}). For the LSST sensitivity limits of $m_i=24, 25, 26$, we find 41, 51, and 60 binaries in our fiducial model. The visualizations of the lensing duration cutoffs are shown for $q=1$ (black) and $q=0.1$ (gray).}
    \label{fig:3-panels}
\end{figure*}

In Figure \ref{fig: PS vs FS}, we also compare the peak magnifications between different wavelengths as a function of orbital inclination. As expected, shorter wavelengths dominate at orbital inclinations close to zero but fall off at longer wavelengths. Thus, longer wavelengths lead to a wider range of orbital inclinations that permit a detectable flare. Between the chosen LSST filter wavelengths, we typically find that discrepancies between $\Delta i_{\rm FS}$ are a few degrees or less, which amount to few $\%$ or less differences in the self-lensing probabilities $P_{\rm PS}, P_{\rm FS}$ and the number of detectable self-lensing flares.

The steep dependence on the minimum lensing duration which occurred for the point-source flares also occurs in the finite-source case. Furthermore, for both point-like and finite sources, the number of detectable flares depends sensitively on the LSST sensitivity limits when no lensing duration was required, but when a minimum lensing duration of 30 days is required, the dependence on the LSST sensitivity limits diminishes. 

Finally, we note that in general, in addition to the finite radial extent of the "minidisks" around the individual SMBHs, one must take into account the finite thickness of these minidisks, as well as the circumbinary gas.  This raises the concern that a thick circumbinary disk for a nearly edge-on binary may obscure the lensing phenomena discussed in our paper.  However, ray-tracing the emission through a hydrodynamical simulation~\cite{Krauth_2024} has shown that the circumbinary disk in the foreground typically does not obscure the self-lensing flare and dip. This is because lensing itself allows us to see into the cavity along bent photon paths.

\section{Summary and conclusions}
\label{sec:DisCon}

In this paper, we estimated the number of binary quasars that are sufficiently edge-on to display detectable self-lensing flares and dips. Our approach relies on the assumption that the same galaxy mergers produce SMBH binaries and activate bright quasars.  Combined with further assumptions about quasar lifetimes, binary fractions, and Eddington ratios, as well as the mass-ratio distributions of binaries extracted from the Illustris simulations, we computed the number of detectable flares, as well as the number of flares with detectable "dips" due to the background SMBH's shadow, in LSST's expected catalog of tens of millions of AGN light-curves. 

We recover earlier results by K21 and find that several hundred lensing flares may be detected by LSST. However, this is based a requiring a minimum flare duration of 30 days. We demonstrated a steep dependence of this number on the minimum required flare duration and found that if much shorter flares (say 10 days) were recoverable, then there would be many more (tens of thousands) of these.

Our main novel result is that we estimate the number of self-lensing dips from SMBHBs. Under plausible assumptions, we find that several dozen of these should be present and detectable by LSST. Finding these dips, a feature of the BHs shadow, would indisputably prove the existence of SMBHBs in distant galaxies, that are currently out of reach for high-resolution VLBI experiments.

\acknowledgments
We thank Luke Krauth for useful discussions. We acknowledge support by NSF grant AST-2006176 and NASA grants 80NSSC22K0822 and 80NSSC24K0440 (ZH). JD acknowledges support by a joint Columbia/Flatiron Postdoctoral Fellowship. Research at the Flatiron Institute is supported by the Simons Foundation. This research was supported in part by the National Science Foundation under Grant No. NSF PHY-1748958 to the Kavli Institute for Theoretical Physics (KITP). This research has made use of NASA's Astrophysics Data System.

{\it Software:} {\tt python} \citep{travis2007,jarrod2011}, {\tt numpy} \citep{walt2011}, {\tt matplotlib} \citep{hunter2007}

\section*{Data Availability}
The data underlying this article will be shared on reasonable request to the corresponding author.

\bibliography{bibfile}

\appendix

\end{document}